\documentclass[conference]{IEEEtran}
\IEEEoverridecommandlockouts   

\usepackage{xcolor}
\usepackage{url}
\usepackage{booktabs}
\usepackage{cite}
\usepackage[hidelinks]{hyperref}
\usepackage{multirow}
\usepackage{array}

\newif\ifshowtodos
\showtodostrue
\showtodostrue % set to \showtodosfalse to hide todos

\usepackage{tikz}
\usetikzlibrary{calc,patterns}
\usetikzlibrary{calc}
\definecolor{axblue}{RGB}{21,101,140}
\definecolor{axred}{RGB}{175,0,30}
\definecolor{axorange}{RGB}{185,70,0}
\begin{document}

\title{Generalization Testing of Grid Foundation Models for Power Flow and Optimal Power Flow}
\author{Ayushi Jolotia and Parikshit Pareek$^{\dagger}$ \\ Indian Institute of Technology Roorkee, India \\ \textit{ 
ayushi\_j@ee.iitr.ac.in, pareek@ee.iitr.ac.in}
\thanks{$^\dagger$Corresponding author.  This work was supported by the ANRF PM Early Career Research Grant (ANRF/ECRG/2024/001962/ENS) and the IIT Roorkee Faculty Initiation Grant (IITR/SRIC/1431/FIG-101078).}}

\maketitle

\begin{abstract}
There is growing interest in foundation models (FMs) for power grids, particularly for power flow and optimal power flow. Their usefulness depends on generalization accuracy, which existing works test on their own benchmarks. We propose a three-axis framework to test their generalization limits: changed admittance matrices of training grids, load levels, and unseen grids. We test GENCO, LUMINA and GridSFM against the DC baseline, alongside their reported limitations, and discuss theoretical reasons for their failures. Experiments reveal that adding a line degrades all FMs more than removing one, and longer lines eventually make LUMINA and fine-tuned GENCO worse than DC. All three degrade beyond their training load range, where the loading-voltage curve steepens. On unseen grids, LUMINA degrades sharply even inside its training range, and GridSFM fails on the grid farthest from it. Thus, beyond training conditions, FMs behave more like meta-models requiring test-time fine-tuning data and compute. 
\end{abstract}


\IEEEpeerreviewmaketitle

\section{Introduction}
\label{sec:intro}
Steady-state assessment of a power grid requires repeatedly solving the power flow (PF) and optimal power flow (OPF) problems. Problems like contingency screening, probabilistic AC-PF and AC-OPF studies under load and renewable uncertainty, and planning across candidate topologies each require thousands of solves~\cite{pfdelta}. Newton–Raphson (NR) for PF and interior-point methods for OPF are reliable, but their cost grows with the number of cases to be solved under uncertainty~\cite{wood2013}. Over the last decade, learned surrogates have been developed for these problems. Surrogates are trained on solver outputs for a large set of inputs and they return a new solution in a single forward pass~\cite{mlopfwiki}. Grid foundation models (FMs) extend this idea using large-scale pretraining and deeper, often attention-based architectures~\cite{genco,gridsfm,hhmpnn,mxgps,unified,lumina}. An FM is trained on a broad dataset and can be adapted to multiple downstream tasks~\cite{bommasani2021}. For a grid FM, this means solving PF or OPF on a network it was never trained on~\cite{genco,gridsfm}. To achieve this, grid FMs are trained on many networks, topologies and loading conditions, and a larger training corpus generally gives better predictions~\cite{genco}. The value of a grid FM therefore rests on the claim that it remains accurate outside its training data, on a new line, a stressed loading or a new network. This property is called generalization. Table~\ref{tab:fms} lists five grid FMs and one closely related model.
\begin{table}[!t]
\vspace{-1.2em}
\renewcommand{\arraystretch}{0.9}
\caption{Foundation and related models for grid assessment}
\label{tab:fms}
\centering\scriptsize
\setlength{\tabcolsep}{3.5pt}
\vspace{-1em}
\begin{tabular}{@{}llll@{}}
\toprule
Model & Task & Pretrained on & Weights \\
\midrule
GENCO / GridFM    & PF, OPF & 100 subgrids, 100--1k bus  & public \\
GridSFM           & AC-OPF  & 54 topologies, 500--4k bus & public \\
HH-MPNN           & AC-OPF  & 14--500 bus grids          & none \\
MxGPS             & PF, SE  & 6 of 8 IEEE grids, 14--300 bus & none \\
LUMINA            & AC-OPF  & 10 grids, 14--13.7k bus    & public \\
Unified grid rep. & PF      & $\le$270 bus encoder       & encoder only \\
\bottomrule
\end{tabular}
\end{table}

\begin{figure}[!t]
\centering
\vspace{-1.3em}
\begin{tikzpicture}[
  bus/.style  = {circle, draw=black!90, fill=white, line width=0.55pt, inner sep=0pt, minimum size=3.0pt},
  nbus/.style = {circle, draw=axblue, fill=axblue, line width=0.45pt, inner sep=0pt, minimum size=3.0pt},
  ln/.style   = {draw=black!85, line width=0.6pt},
  gone/.style = {draw=axred, line width=0.5pt, dash pattern=on 2.0pt off 1.6pt},
  new/.style  = {draw=axblue, line width=0.85pt},
  head/.style = {font=\footnotesize\bfseries, align=center},
  what/.style = {font=\footnotesize, text=black, align=center, text width=2.75cm},
  tag/.style  = {font=\fontsize{7pt}{7.8pt}\selectfont, inner sep=1pt},
  sep/.style  = {draw=black!25, line width=0.4pt}
]
\def\PW{2.92}
\foreach \i/\head/\what in {%
  0/{A1\\Admittance matrix}/{A1.1: line \textcolor{axred}{out} or \textcolor{axblue}{added}
  A1.2: \textcolor{axorange}{longer lines}},
  1/{A2\\Loading region}/{Load beyond the training range of FM},
  2/{A3\\Unseen grid}/{A network not in the training set}}
{
  \node[head] at (\i*\PW+1.40,3.06) {\head};
  \draw[axblue,line width=0.8pt] (\i*\PW+0.25,2.74) -- (\i*\PW+2.55,2.74);
  \node[what,anchor=north] at (\i*\PW+1.40,0.10) {\what};
  \ifnum\i<2 \draw[sep] (\i*\PW+2.80,-0.45) -- (\i*\PW+2.80,3.22);\fi
}

\begin{scope}[shift={(1.40,1.55)},scale=0.78]
  \node[bus](p0) at (-0.965, 0.800){};  \node[bus](p1) at (-0.125, 0.352){};
  \node[bus](p2) at ( 0.570, 0.800){};  \node[bus](p3) at (-1.100,-0.032){};
  \node[bus](p4) at ( 0.236,-0.050){};  \node[bus](p5) at (-0.410,-0.690){};
  \node[bus](p6) at ( 1.100,-0.524){};
  \draw[ln] (p0)--(p1) (p1)--(p3) (p1)--(p2) (p3)--(p4) (p1)--(p5) (p3)--(p5) (p2)--(p4) (p2)--(p6) (p4)--(p6);
  \draw[gone] (p0)--(p3);
  \coordinate (x) at ($(p0)!0.5!(p3)$);
  \draw[axred,line width=0.8pt] ($(x)+(-0.08,-0.08)$) -- ($(x)+(0.08,0.08)$) ($(x)+(-0.08,0.08)$) -- ($(x)+(0.08,-0.08)$);
  \draw[new] (p5)--(p6);
  \node[nbus](q) at (1.15,0.75){}; \draw[new] (p2)--(q);
  \draw[axorange,line width=1.4pt] (p1)--(p5);
  \node[font=\footnotesize,text=axorange,anchor=north,inner sep=1pt] at ($(p5)+(0.02,-0.08)$) {$\times m$};
\end{scope}

\begin{scope}[shift={(\PW+0.42,1.52)}]
  \fill[black!15] (0,0) rectangle (0.95,1.08);
  \draw[->,black!85,line width=0.5pt] (0,0) -- (2.15,0);
  \draw[->,black!85,line width=0.5pt] (0,0) -- (0,1.16);
  \draw[black,line width=0.9pt] plot[domain=0:1,samples=80] ({1.90*\x},{0.50+0.50*sqrt(1-\x)});
  \draw[black!65,line width=0.6pt,dash pattern=on 1.8pt off 1.5pt] plot[domain=0:1,samples=80] ({1.90*\x},{0.50-0.357*sqrt(1-\x)});
  \draw[axred,line width=1.2pt,dash pattern=on 3.0pt off 1.6pt] (0.95,0.854) -- (2.08,0.644);
  \draw[densely dotted,black!75] (1.90,0) -- (1.90,0.50);
  \fill[black] (1.90,0.50) circle (1.3pt);
  \node[tag,text=axred,anchor=south] at (1.40,0.84) {FM};
  \node[tag,anchor=north east] at (1.52,0.77) {True};
  \node[tag,anchor=east] at (-0.02,0.96) {$|V|$};
\node[tag,text=black!80,align=center] at (0.47,0.46) {Training\\[-3pt]range};
  \node[tag,anchor=north east] at (1.97,0.31) {$\lambda^\star$};
  \node[tag,anchor=north east] at (2.15,-0.02) {Loading Level};
\end{scope}

\begin{scope}[shift={(\PW+0.42,0.40)}]
  \fill[black!15] (0,0) rectangle (0.95,0.66);
  \draw[->,black!85,line width=0.5pt] (0,0) -- (2.15,0);
  \draw[->,black!85,line width=0.5pt] (0,0) -- (0,0.74);
  \draw[axred,line width=1.1pt] plot[domain=0:1.89,samples=60] (\x,{0.05+0.58*(max(\x-0.95,0)/0.95)^2.4});
  \draw[densely dotted,black!75] (1.90,0) -- (1.90,0.66);
  \node[tag,anchor=south,rotate=90,text=axred] at (-0.09,0.42) {Error};
  \node[tag,anchor=north east] at (2.15,-0.02) {Loading Level};
\end{scope}

\begin{scope}[shift={(2*\PW+1.40,1.55)},scale=0.82]
  \begin{scope}[shift={(-0.05,0.8)}]
    \node[bus](s0) at (-0.148, 0.335){};  \node[bus](s1) at ( 0.435, 0.233){};
    \node[bus](s2) at (-0.300,-0.040){};  \node[bus](s3) at ( 0.600,-0.535){};
    \node[bus](s4) at (0.10,-0.535){};
    \draw[ln] (s0)--(s1) (s0)--(s2) (s1)--(s2) (s0)--(s3) (s1)--(s3) (s1)--(s4) (s2)--(s4);
    \node[tag,text=black!85] at (-0.95,-0.36) {Trained on};
  \end{scope}
  \node[font=\scriptsize,text=red!95] at (-0.05,-0.05) {\large $\downarrow$};
  \begin{scope}[shift={(0.05,-0.8)},scale=0.85]
    \node[bus](A) at (-0.52, 0.546){};  \node[bus](B) at (-0.09, 0.240){};
    \node[bus](C) at ( 0.39, 0.546){};  \node[bus](D) at (-0.75,-0.022){};
    \node[bus](E) at ( 0.16,-0.109){};  \node[bus](F) at (-0.28,-0.470){};
    \node[bus](G) at ( 0.75,-0.153){};
    \draw[ln] (A)--(B) (A)--(D) (B)--(D) (B)--(C) (D)--(E) (B)--(F) (D)--(F) (C)--(E) (E)--(G) (C)--(G);
    \node[tag,text=axblue] at (-1.35,-0.38) {Applied to};
  \end{scope}
\end{scope}
\end{tikzpicture}
\vspace{-23pt}
\caption{Proposed generalization axes. Left: A1, an unseen $\rm Y_{\mathrm{bus}}$ (changed incidence matrix or condition number) of a pretraining network. Center: A2, load beyond the training range. Right: A3, a network not seen in pretraining.}
\label{fig:axes}
\vspace{-1em}
\end{figure}
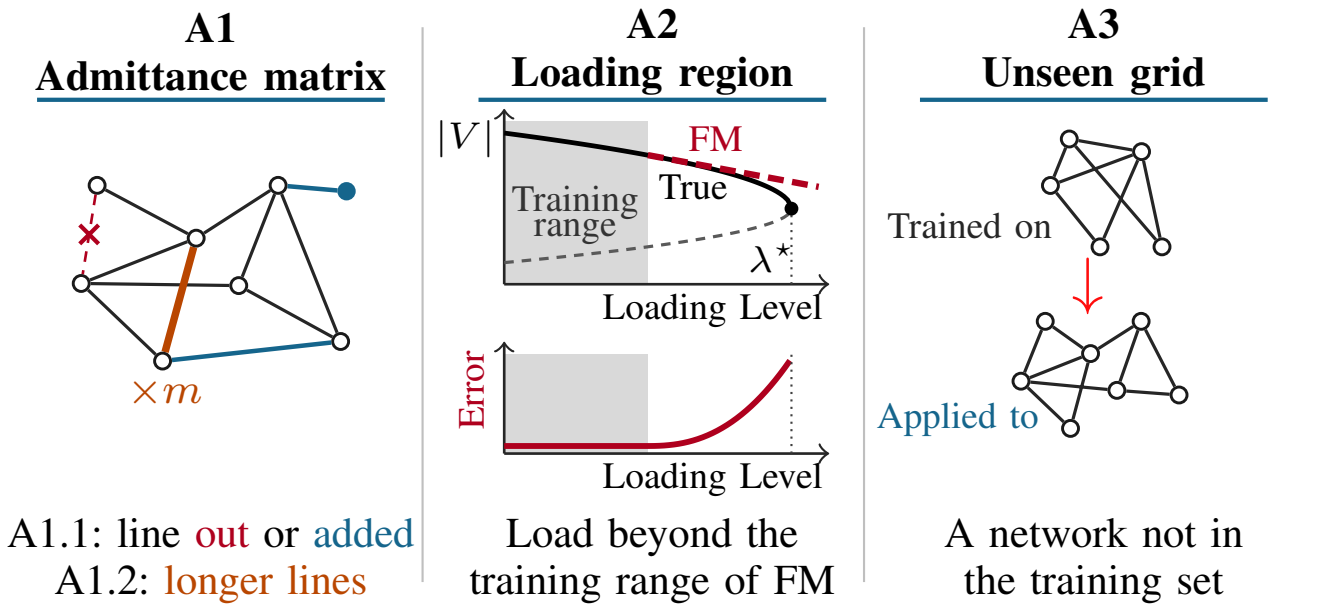

Importantly, the FMs' proposed use cases are the main motivation for testing the generalization claim. An FM can be used in a \textit{zero-shot} manner, where a test case is predicted directly with frozen weights, without any solved cases from the target network. An FM can also be used after \textit{fine-tuning}, with its weights updated on solved cases of the target network. Fine-tuning recovers accuracy but requires solved-instance data and training compute resources for each target network. For example, GridSFM was fine-tuned with 100 solved cases on a 10,000-bus grid, and HH-MPNN with 15k on a 2000-bus grid~\cite{gridsfm,hhmpnn}. Further, an FM can be used to \textit{warm-start} an exact solver. In that case, the FM acts as a `learned initializer', and the convergence guarantee comes from the solver, which either converges or provides a failure certificate. Here, we test the limits of FMs used in the first two modes, i.e., zero-shot or fine-tuned predictors. In these modes, an FM returns a single answer, without any certificate of failure or error estimate, so its failures are silent. Therefore, testing an FM's limits matters more than its average accuracy.

In this paper, we propose a three-axis framework for testing the generalization of grid FMs (Fig.~\ref{fig:axes}). Each axis comes with a measure of how far a test case lies from the training data in terms of either the grid or bus-injection data. The first axis (\textbf{A1}) keeps a power network the model was trained on but changes its admittance matrix $\rm Y_{\mathrm{bus}}$ by removing or adding lines, or lengthening transmission lines. On this axis, the distance from the training data is the size of the change, measured by the severity of a graph change or by the length multiplier  $m$. The second axis (\textbf{A2}) keeps a training grid fixed and raises the load beyond the range seen in training, up to and past the loading limit. Its distance is the load as a fraction of that limit. The third axis (\textbf{A3}) applies the model to a network it has never seen. Its distance is judged from grid properties such as condition number, $R/X$ ratio, loops per bus and depth, compared with the range covered by the training grids. We note that the existing literature covers these axes only in parts and sporadically, each work with its own metrics. To our knowledge, no grid FM has been tested on added components, since the data generators behind them provide at most $\rm N$-$\rm k$ conditions. We apply the proposed three-axis framework to three released models, GENCO~\cite{genco} for AC-PF, and LUMINA~\cite{lumina} and GridSFM~\cite{gridsfm} for AC-OPF. The ground truth is computed with PowerModels~\cite{powermodels}, using NR for AC-PF and Ipopt~\cite{wachter2006ipopt} for AC-OPF, while DC-PF or DC-OPF on the same samples gives a baseline that a useful model should beat. For GENCO, we compare the pretrained zero-shot checkpoint with the available checkpoints fine-tuned on 1k to 250k solved cases of the 118-bus grid to highlight the effects of compute and data. Our main contributions are:

\begin{itemize}
    \item Designing a three-axis framework for testing generalization limits of grid FMs, with a measure of distance from the training data on each axis (Fig.~\ref{fig:axes}). We further provide theoretical arguments as possible explanations of the failures on each axis.

    \item Testing three released FMs, GENCO, LUMINA and GridSFM, along all three axes. On \textbf{A1}, the admittance matrix axis, we find that line addition is more severe than line outage, and increasing line length eventually makes LUMINA and even fine-tuned GENCO worse than DC approximations. Further, on \textbf{A2}, the loading axis, all FMs degrade beyond their training load range, where the $P$--$V$ curve steepens. Finally, on \textbf{A3}, the unseen-grid axis, the FMs' zero-shot performance on new networks is worse than that of DC approximations, and an FM may fail even on an unseen grid whose properties lie within the range of its training grids.

\end{itemize}
The following sections cover A1 to A3 in order, followed by the discussion  and conclusion section.

\section{A1: Admittance Matrix Axis}
\label{sec:admittance}
The  $\rm Y_{\mathrm{bus}}$ of a network encodes its topology and line parameters. The training data of the three FMs at most remove a line and vary line parameters within $\pm20\%$, and the published tests only remove or switch existing lines, with errors rising from 1.05$\times$ to $69\times$ over the training networks~\cite{hhmpnn,liu2023topologyaware}. We therefore propose two tests beyond these: A1.1) network graph changes (Section~\ref{sec:graph}) and A1.2) longer lines (Section~\ref{sec:line}).

\subsection{Graph Modifications: \texorpdfstring{$\rm N$-$\rm k$ and $\rm N$+$\rm 1$}{N-k and N+1}}
\label{sec:graph}
In this sub-axis, we change the network graph by removing lines ($\rm N$-$\rm k$) or adding one line ($\rm N$+$\rm 1$), while keeping the base load constant to isolate the effect of the change on $\rm Y_{\mathrm{bus}}$. Such changes occur in practice through maintenance, faults and network expansion, and the topology given to a model can also be wrong due to topology estimation errors~\cite{abur2004}. A removal takes out a single line or three to five lines ($\rm N$-$ \rm k$), and an addition adds a parallel line or a new line between two buses that are two lines apart (short new line) or seven or more lines apart (long new line). A short new line has the grid's median line parameters, and a long new line has these values doubled. The severity of a change (adding or removing a line) is defined as the ratio of the power flow on that line to its rating, measured before and after the change. To test medium and severe changes of each type, we keep cases leading to severities of 0.6 and 0.9, such that the modified grid stays within its voltage limits and has an AC-OPF solution. Note that three to five lines out are built only at severity 0.6, and the parallel line takes about half the flow of the line it connects, so it never reaches severity 0.9. We test on the 118- and 2000-bus grids, which the models (except zero-shot GENCO) were trained on, drawing 1k and 500 new samples, respectively, with each FM's own data generator.

\begin{figure}[t]\centering

\includegraphics[width=\columnwidth]{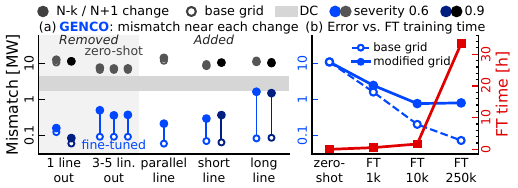}
\vspace{-2em}
\caption{GENCO on 118-bus. (a) Mean real power mismatch over the buses within three hops of each removed or added line, base grid (hollow) and modified grid (filled). (b) The same mismatch averaged over line-addition cases, with fine-tuning (FT) training time (red) from the authors' logs.}
\label{fig:a11_case118_genco}
\vspace{-1em}
\end{figure}

Considering locality of voltage and reactive power in transmission grids, we measure mismatch at buses within three hops of a removed or added line for GENCO, and at the three nearest generator buses for the GridSFM performance measure. Further, LUMINA's dispatch error is measured over all generators, since the active-power changes are global. Fig.~\ref{fig:a11_case118_genco}a shows that zero-shot GENCO is worse than DC-PF in every modified-grid case, which is consistent with reporting in~\cite{genco}. In contrast, fine-tuned (FT) GENCO's error rises up to $5.5\times$ after $\rm N$-$\rm k$ and up to $20\times$ after $\rm N$+$\rm 1$, while staying below DC-PF in all cases. However, this improvement comes at the cost of solved samples and training time: Fig.~\ref{fig:a11_case118_genco}b shows that the error on the base grid decreases steadily with both, but on the modified grid it saturates at 10k samples, and at 250k samples it remains more than 10 times that on the base grid.

\begin{figure}[!t]\centering
\vspace{-0.1em}
\includegraphics[width=\columnwidth]{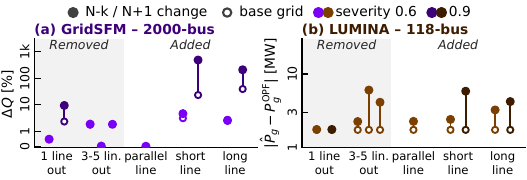}
\vspace{-2.1em}
\caption{(a) GridSFM on 2000-bus: reactive violation $\Delta Q$, mean over the three generator buses nearest to the topology change, over 500 samples. (b) LUMINA on 118-bus: dispatch error, mean over all generators in each sample, median over 1k samples. Base grid (hollow), modified grid (filled).}

\label{fig:a11_case2000}
\end{figure}

Next, we use GridSFM to predict generator set-points for the 2000-bus grid and solve an AC-PF from them via NR, following~\cite{gridsfm}, without PV--PQ bus conversion at reactive limits. Therefore, we measure the generators' reactive-limit violation $\Delta Q$ as a percentage of $Q_{\max}-Q_{\min}$. We observed that, on the base grid, $\Delta Q$ is $7\times$ lower than with DC-OPF set-points, it stays below 10\% after lines are removed, and rises from 23 to 482\% after a short new line is added at severity 0.9 (Fig.~\ref{fig:a11_case2000}a). Importantly, the generation-cost error reported in~\cite{gridsfm} stays at 1.1\% and misses this reactive-limit failure. Further, for LUMINA on the 118-bus grid, Fig.~\ref{fig:a11_case2000}b shows that the dispatch error rises up to $3.5\times$ after four lines out, and reaches DC-OPF's after new lines are added at severity 0.9.

\textbf{Theoretical Discussion:}
Numerical results clearly indicate that line addition moves the test system beyond the $\rm N$-$\rm k$ based training set, i.e., outside the convex hull of an outage-based training set. This is because adding a line to the reduced susceptance matrix $B_0\succ0$ gives $B_+=B_0+baa^\top$, where $b>0$ and $a$ is the reduced line-incidence vector~\cite{dorfler2013kron}. Addition moves eigenvalues away from zero, whereas outages move them towards zero (some eigenvalues remain unchanged). For a fixed base grid with positive reactances and unchanged remaining line parameters, each outage matrix satisfies $B_D\preceq B_0$. This inequality is also preserved by convex combinations, so $B_+$ lies outside the convex hull of these outage matrices. Similarly, GENCO's training data have only one line out, so three to five lines out are also new to it, and its error rises more after them than after one line out. LUMINA and GridSFM were trained only on intact grids, so every change, including removal, is new to them. Thus, outage-only coverage leaves an extrapolation requirement, which can challenge learned models~\cite{xu2021extrapolate}. We also note that this DC-based analysis is only indicative of the AC-PF and AC-OPF performance.

\subsection{Line Length Modifications: \texorpdfstring{$(R, X, b_{\mathrm{ch}})\times m$}{(R, X, bch)x m}}
\label{sec:line}
\begin{figure}[!t]\centering
\vspace{-0.2em}
\includegraphics[width=\columnwidth]{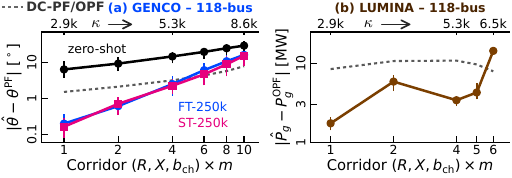}
\vspace{-2.2em}
\caption{Longer corridor on 118-bus; $m=1$ is the unchanged grid. (a) GENCO's mean angle error over the buses within three hops of the corridor, median over the 531 samples with a true PF solution at every $m$ (ST: scratch-trained). (b) LUMINA's dispatch error, mean over all generators, median over 1k samples. Bars: 25th to 75th percentile.}
\label{fig:a12_case118}
\end{figure}

In this test, we increase the length of transmission lines $m$ times by multiplying line parameters (resistance, reactance, and shunt) by $m$, which is analogous in practice to line parameters in the network model being off, or to long tie-lines between areas weakening the grid. Note that our test keeps the $R/X$ ratio constant while raising the condition number $\kappa$ (the ratio of the largest to smallest eigenvalue) of the susceptance matrix $B$. A higher $\kappa$ means the matrix is closer to singular, so small changes in power injections can cause large changes in bus angles. To raise $\kappa$, we lengthen the lines contributing most to the smallest eigenvalue of $B$: 12 lines in the 118-bus grid and 20 in the 2000-bus grid. This makes $\kappa$ rise from 2.9k to 8.6k at $m=10$ in the 118-bus grid and from 193k to 545k on the 2000-bus grid at $m=8$. Note that GENCO's training data vary each line parameter by $\pm20\%$, and LUMINA's and GridSFM's keep the values of the original case files.  
We plot GENCO's angle error within three hops of the lengthened corridor. Note that a line $m$ times longer carries $m$ times less power for the same angle error. Thus, angle error presents a fair measure of performance for this axis. Fig.~\ref{fig:a12_case118}a shows that a fine-tuned GENCO is $7.6\times$ more accurate than DC-PF at training length, as good as DC-PF at $m\approx5$, and performs $>2\times$ worse at $m=10$, with an error of 16.3$^\circ$. Also, a scratch-trained model has similar performance, while the zero-shot GENCO is worse than DC-PF at every length.

\begin{figure}[!t]\centering
\vspace{-1em}
\includegraphics[width=\columnwidth]{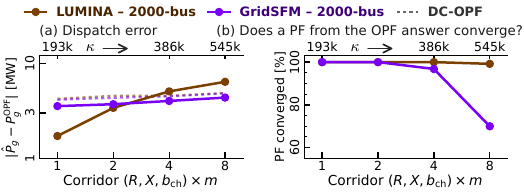}
\vspace{-2.2em}
\caption{Longer corridor on 2000-bus, as in Fig.~\ref{fig:a12_case118}, 500 samples per model. (a) Dispatch error, mean over all generators, median over samples; dotted: DC-OPF on each model's samples. (b) Share of samples whose AC-PF from the model's set-points converges.}
\label{fig:a12_case2000}
\vspace{-1em}
\end{figure}

For LUMINA and GridSFM, we plot the dispatch error over all generators. Note that LUMINA's inputs do not include a line's $R$, $X$ or rating, so its answer stays the same at every $m$, while the AC-OPF dispatch moves. Fig.~\ref{fig:a12_case118}b shows that LUMINA is $5\times$ more accurate than DC-OPF at training length and $1.8\times$ worse at $m=6$, with an error of 14.5\,MW, and on the 2000-bus grid it is worse than DC-OPF from $m=4$ (Fig.~\ref{fig:a12_case2000}a). Importantly, its own PF mismatch (can be checked without true answer) stays at 136MW on the 118-bus grid at every $m$. Thus, it fails without error growth warning. In contrast, GridSFM's inputs include the line parameters, and its dispatch error is $\leq$ DC-OPF's at every $m$, while the AC-PF from its prediction converges for only 70\% of samples at $m=8$ (Fig.~\ref{fig:a12_case2000}b).

\textbf{Theoretical Discussion:} Unlike an outage, an increase in line length only reduces its flow. A common length multiplier $m>1$ preserves $R/X$ but reduces each selected line's DC weight from $b_e = 1/x_e$ to $b_e/m$. Hence, $B(m) = B_0 - (1-m^{-1})\sum_{e\in S} b_e a_e a_e^{\top} = m^{-1}B_0 + (1-m^{-1})B_{\mathrm{out},S}$,
where $B_{\mathrm{out},S}$ is the matrix with all selected lines removed~\cite{dorfler2013kron}. With higher $m$ the matrix approaches a joint line outage, but in a smooth manner unlike graph changes. The training data, however, contain outages of at most $k$ lines, and $B(m)$ lies in the convex hull of $B_0$ and these outage matrices only if $|S|(1-m^{-1})\le k$, i.e., $m \le |S|/(|S|-k)$ for $k<|S|$. With $|S|=12$ and $20$ lines lengthened here, every tested $m\ge 2$ lies outside the hull whenever $k<|S|/2$, and thus demands extrapolation from the FMs. For the single-line outages ($k=1$) in GENCO's training data, the bound is $m\le 1.09$.

\section{A2: Loading Region Axis}
\label{sec:load}
As the load increases, the PF problem has no solution beyond a maximum-loading point $\lambda^\star$~\cite{ajjarapu1992}, and an AC-OPF can fail even earlier once the generators reach their limits. The training data of the three FMs scale the total load to about 1.25$\times$ the base load, and the only published test beyond this range reports an error 81.5$\times$ larger at the loading limit for GENCO~\cite{genco}. We raise the load from within the training range up to and beyond $\lambda^\star$ (for LUMINA and GridSFM, the largest $\lambda$ with an AC-OPF solution) to test how an FM's error grows.
The loads of each test sample are scaled to the total load of the base grid to keep each sample's load pattern intact as we increase the load factor $\lambda$ in steps of $\lambda/\lambda^\star$. For GENCO on the 118-bus grid, $\lambda^\star$ lies at 2.45 to 3.43$\times$ the base load, and for LUMINA and GridSFM on the 2000-bus grid at 1.31$\times$.

\begin{figure}[!t]\centering
\vspace{-0.5em}
\includegraphics[width=\columnwidth]{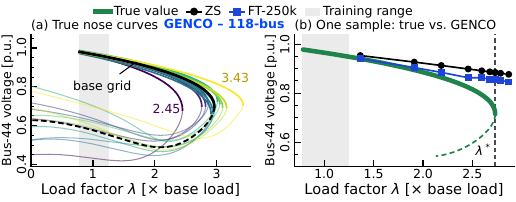}
\vspace{-2.2em}
\caption{GENCO on 118-bus. (a) True bus-44 voltage against load for 12 samples and the base grid. (b) One sample: true voltage and GENCO's answers; dashed line, the loading limit $\lambda^\star$. Grey: training range.}

\label{fig:a2_nose}
\end{figure}

\begin{figure}[!t]\centering
\vspace{-0.4em}
\includegraphics[width=\columnwidth]{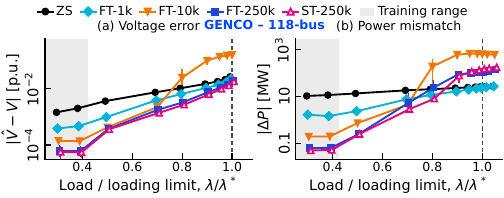}
\vspace{-2.2em}
\caption{GENCO on 118-bus, against $\lambda/\lambda^\star$, each sample's load as a fraction of its loading limit. (a) Voltage error, mean over the 64 load buses. (b) Power mismatch, mean over buses. Median over 990 samples, bars 25 to 75\%.}
\label{fig:a2_genco}
\end{figure}

We plot the voltage at bus 44 of the 118-bus grid, which has the lowest voltage and falls most steeply near $\lambda^\star$ in 969 of 1k test samples, making it the hardest one to follow. Fig.~\ref{fig:a2_nose}a shows the true nose curves of bus 44 for 12 of GENCO's test samples, traced through $\lambda^\star$ with a continuation power flow~\cite{ajjarapu1992}. Fig.~\ref{fig:a2_nose}b shows that FT GENCO follows the true voltage inside the training range, but beyond it its voltage falls much slower than the true one, and its voltage error and mismatch grow by two to three orders of magnitude at $\lambda^\star$ (Fig.~\ref{fig:a2_genco}). Zero-shot GENCO performs poorly throughout, and the 10k FT model breaks down from $\lambda/\lambda^\star=0.8$.

\begin{figure}[!t]\centering
\vspace{-0.35em}
\includegraphics[width=\columnwidth]{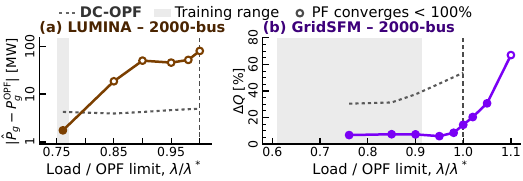}
\vspace{-2em}
\caption{LUMINA and GridSFM on 2000-bus, against $\lambda/\lambda^\star$, each sample's load as a fraction of its OPF limit. (a) LUMINA's dispatch error. (b) GridSFM's reactive violation $\Delta Q$ over all generators.}
\label{fig:a2_case2000}
\end{figure}

Fig.~\ref{fig:a2_case2000}a shows LUMINA's dispatch error on the 2000-bus grid is $2.5\times$ lower than DC-OPF's at the base load, but reaches $16\times$ DC-OPF's at $\lambda^\star$, and from $\lambda/\lambda^\star=0.90$ no PF from LUMINA's answer converges. In contrast, GridSFM's dispatch error and $\Delta Q$ stay below DC-OPF's, with $\Delta Q$ doubling to 15\%, and a PF from its answer converges up to $\lambda^\star$ (Fig.~\ref{fig:a2_case2000}b). Even past $\lambda^\star$, GridSFM still answers: at $\lambda/\lambda^\star=1.10$, its $\Delta Q$ reaches 67\%, and a PF from 40\% of its answers does not converge. Both OPF models share the same $\lambda^\star$, but their training loads reach only $1.01\times$ the base load for LUMINA and $1.20\times$ for GridSFM, which explains why GridSFM error growth is slower compared to LUMINA.

\textbf{Theoretical Discussion:} It is well established that as loading increases, the $P$--$V$ curve becomes steeper near its nose, so a small increase in demand can produce a large voltage change~\cite{ajjarapu1992}. Learned models and FMs trained mainly around normal loading fit the flatter part of this curve but fail to capture its changing slope outside the training range~\cite{xu2021extrapolate}. This is an extrapolation failure: a PF solver corrects the state using the nonlinear equations at each loading level and checks convergence, whereas an FM relies on its training data to represent the variable nonlinearity.

\section{A3: Unseen Grid Axis}
\label{sec:unseen}
\begin{figure}[!t]
    \centering
 \vspace{-0.7em}
    \includegraphics[width=\linewidth]{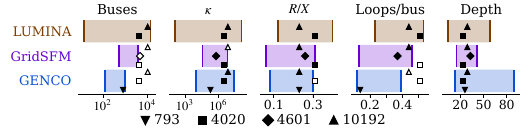}
    \vspace{-2.3em}
\caption{Training range of each model's grid properties (bands) and the unseen test grids it was run on (markers; filled: inside the range, hollow: outside). $R/X$ is the branch median, and depth is in hops from the slack.}

\label{fig:grid_measures}
\vspace{-1em}
\end{figure}

This axis is generally considered the hardest as it clearly demands generalization from learned solvers like FMs. Published tests on unseen grids report errors up to 3222$\times$ their in-distribution values, two models less accurate than DC-PF, and an AC-OPF warm-started from GridSFM's answer that never converges on a 10,000-bus grid~\cite{genco,lumina,gridsfm,hhmpnn,lopez2023typedgnn,mxgps}. In this test, we run each FM zero-shot on PGLib grids~\cite{pglib} outside its training set, as a utility would when applying a released FM to its own network. The samples are drawn by each model's own data generator, so the only variable is the test grid. Of 500 samples drawn per unseen grid, those whose true solution could be computed are kept (302 to 500). Each unseen grid is compared with each FM's training range on five grid properties, namely, the number of buses, condition number $\kappa$, $R/X$, loops per bus, and depth from the slack (Fig.~\ref{fig:grid_measures}). Each unseen grid is then classified as: near if all five lie within the FM's training range, edge if any lies at most 25\% outside it, and far if any lies further outside or it comes from a grid family not in training (Table~\ref{tab:a3}). We compute each FM's error using the metrics reported in its paper where available~\cite{genco,gridsfm}, and express it relative to DC-PF/DC-OPF on the same samples ($\times$DC). Table~\ref{tab:a3} shows that zero-shot GENCO's mismatch error is worse than DC-PF's on every grid including one of the training grids, which is consistent with the reporting in~\cite{genco}. In contrast, LUMINA is 5$\times$ more accurate than DC-OPF in dispatch on its training grid, despite a high mismatch error, but 17$\times$ worse on the near grid, where the PF from its answer converges for only 22\% of samples. For GridSFM, the dispatch error stays close to its training-grid value on the edge grids, but grows to 10$\times$ DC-OPF's on the far grid.
So, the generalization capacity of FMs on a new, unseen grid is not decided just by its closeness to the training grids in the five grid properties above; rather, it depends on many factors, including but not limited to these properties. They therefore give an idea, but no guarantee, of how far an FM can generalize to the grid. In short, the generalization of FMs on unseen grids does not follow a simple trend.

\textbf{Theoretical Discussion:} Under the DC approximation, the bus angles follow $\theta=B^{-1}p$, where $B$ is the grid's reduced susceptance matrix~\cite{dorfler2013kron}.
The inverse depends on network connectivity and line parameters, so changes to the grid change the solution map, i.e., how injections influence states across the network. Similar network sizes or condition numbers do not determine this map, since they do not specify the full spectrum and eigenvectors of $B$ or of the PF Jacobian. A shared graph architecture can process the new network, but its learned operations need not approximate the new solution map. Accurate transfer therefore depends on learning or encoding a solution rule that remains valid across different networks~\cite{xu2021extrapolate}, as equation-based solvers do and FMs appear not to.

\vspace{-0.7em}
\begin{table}[!t]
\renewcommand{\arraystretch}{0.75}
\caption{Model performance on seen (training) and unseen grids, relative to DC (PF/OPF).}
\label{tab:a3}
\centering\footnotesize\setlength{\tabcolsep}{1.2pt}
\vspace{-1em}
\begin{tabular}{@{}llcccccc@{}}
\toprule
\multicolumn{2}{@{}l}{\textbf{GENCO} \textit{(PF model)}} & \multicolumn{2}{c}{Mismatch [MW]} & \multicolumn{2}{c}{Angle [$^\circ$]} & \multicolumn{1}{c}{$|V|$ [$10^{-3}$]} \\
\cmidrule(lr){3-4}\cmidrule(lr){5-6}\cmidrule(lr){7-7}
Grid & Type & Model & $\times$DC & Model & $\times$DC & \multicolumn{1}{c}{Model} \\
\cmidrule{1-8}
Subgrid$^a$ & Seen & 6.44 & 24 & 7.09 & 2.7 & \multicolumn{1}{c}{11.8} \\
793 & Near & 19.5 & 14 & 6.18 & 1.9 & \multicolumn{1}{c}{9.36} \\
4020 & Far & 37.6 & 8.6 & 11.8 & 7.8 & \multicolumn{1}{c}{5.92} \\
10192 & Far & 22.0 & 72 & 4.33 & 0.5 & \multicolumn{1}{c}{14.6} \\
\midrule
\multicolumn{2}{@{}l}{\textbf{GridSFM} \textit{(OPF model)}} & \multicolumn{2}{c}{$P_g$ MAPE [\%]} & \multicolumn{2}{c}{PF angle [$^\circ$]} & \multicolumn{1}{c}{$|V_{\mathrm{gen}}|$ [$10^{-3}$]} \\
\cmidrule(lr){3-4}\cmidrule(lr){5-6}\cmidrule(lr){7-7}
Grid & Type & Model & $\times$DC & Model & $\times$DC & \multicolumn{1}{c}{Model} \\
\cmidrule{1-8}
500 & Seen & 3.83 & 1.6 & 2.74 & 1.1 & \multicolumn{1}{c}{7.04} \\
4020 & Edge & 2.28 & 1.5 & 2.89 & 1.3 & \multicolumn{1}{c}{5.54} \\
4601 & Edge & 3.05 & 2.2 & 2.46 & 3.0 & \multicolumn{1}{c}{7.22} \\
10192 & Far & 28.5 & 10 & 26.3 & 24 & \multicolumn{1}{c}{50.8} \\
\midrule
\multicolumn{2}{@{}l}{\textbf{LUMINA} \textit{(OPF model)}} & \multicolumn{2}{c}{Mismatch [MW]} & \multicolumn{2}{c}{Angle [$^\circ$]} & \multicolumn{2}{c}{$P_g$ [MW]} \\
\cmidrule(lr){3-4}\cmidrule(lr){5-6}\cmidrule(lr){7-8}
Grid & Type & Model & $\times$DC & Model & $\times$DC & Model & $\times$DC \\
\cmidrule{1-8}
118 & Seen & 136 & 45 & 4.20 & 1.0 & 1.83 & 0.2 \\
4020 & Near & 9.6k & 2.2k & 10.2 & 3.8 & 30.7 & 17 \\
10192 & Far & \multicolumn{2}{c}{Diverged}  & \multicolumn{2}{c}{Diverged} & \multicolumn{2}{c}{Diverged} \\
\bottomrule
\end{tabular}
\par\vspace{2pt}\noindent{\scriptsize
 All values are mean absolute errors (MAE), except GridSFM's $P_g$ mean absolute percentage error (MAPE). $^a$A training subgrid part of the 19,402-bus grid.\par}
\end{table}

\section{Discussion and Conclusion}
\label{sec:DC}
We explored the generalization limits of grid FMs from both theoretical and empirical perspectives, including results reported in the literature, within our proposed three-axis framework. The three axes of the framework are changed admittance matrices of training grids, loading beyond the training range, and unseen grids. Our analysis indicates that current FMs perform well near the conditions covered by their training data and by what their inputs encode (as in Sections~\ref{sec:admittance} and~\ref{sec:load}), but do not generalize well beyond them. Further, they fail without any indication (silently). Reliable verification of the point estimates of the FMs then requires extra solves from the solver, and improving the performance of the model on the test network requires \textit{solved samples and training time} to fine-tune the model. This increases the test-time cost (GENCO in Section~\ref{sec:admittance}) and reduces the advantage of using FMs over learned surrogates. On an unseen grid, the generalization ability of FMs does not follow a simple trend. This is because the grid properties used to judge the closeness of an unseen grid to the training data do not fully determine the unseen grid's solution map (Section~\ref{sec:unseen}), so there is no guarantee of the model's accuracy even if the new grid lies within the training range, as seen with LUMINA. Overall, our analysis suggests that, beyond their training grids, current FMs behave like meta-models, which require both test-time compute and solved samples to enhance their performance on unseen grids.


\vspace{-1em}

\end{document}